\documentclass[useAMS,usenatbib]{mn2e}
\usepackage{graphicx}
\usepackage{amsmath,amssymb}

\def\note #1]{\noindent{\bf #1]}}

\newcommand{\D}{\displaystyle}

\begin{document}

\title[Pulsations in magnetic Ap stars]{Improved pulsating models of  magnetic Ap stars I: exploring different magnetic field configurations}

\author[M.S. Cunha]
{ M.S. Cunha\\ 
Centro de Astrof\'\i sica da Universidade do Porto, Rua das
 Estrelas, 4150 Porto, Portugal}

%\date{Accepted 2003;  Received 2003; in original form 2003 March 18} 
\pubyear{2006}

\maketitle
%-----------------------------------------------------------------

\begin{abstract}

Magnetic perturbations to the frequencies of low degree, high radial order, axisymmetric pulsations in stellar models permeated by large scale magnetic fields are presented. Magnetic fields with dipolar, quadrupolar and a superposition of aligned dipolar and quadrupolar components are considered. The results confirm that the magnetic field may produce strong anomalies in the power spectra of roAp stars. It is shown for the first time that anomalies may result both from a sudden decrease or a sudden increase of a mode frequency. Moreover, the results indicate that the anomalies depend essentially on the geometry of the problem, i.\ e.\, on the configuration of the magnetic field and on the degree of the modes. This dependence opens the possibility of using these anomalies as a tool to learn about the magnetic field configuration in the magnetic boundary layer of pulsating stars permeated by large scale magnetic fields.  

\end{abstract}


\section{Introduction}
Rapid oscillations, with amplitudes of a few mmag have first been discovered in cool, chemically peculiar magnetic Ap stars over two decades ago by \citet{kurtz82}. The frequencies, $\nu$, of the observed oscillations are typically within the range $1$mHz $<\nu< 3$mHz, although lower frequency pulsations have been predicted by \citet{cunha02} - and recently found by \citet{elkin05} -  among the more evolved cool Ap stars. Recent observational and theoretical reviews this class of  pulsators, known as roAp stars, are provided in \citet{kurtz00} and \citet{kurtz04}, and \citet{cunha03,cunha05} and \citet{gough05}.

The oscillations observed in roAp stars are believed to be high order acoustic modes modified near the surface by the magnetic field. In principle, the high order of the observed oscillations should simplify the theoretical analysis of the pulsation power spectra of roAp stars, by allowing the use of standard asymptotic tools. For this reason, roAp stars have been recognized as excellent potential targets for asteroseismology. However, as emphasized in the works of \citet{matthews99} and \citet{cunhaetal03}, the pulsation power spectra of these stars cannot be reconciled with that produced with standard stellar and pulsating models. Part of the discrepancy may have its origin in the difficulty of deriving accurate effective temperatures and bolometric corrections (and hence radii) for Ap stars \citep{leckrone73,stepien94}. However, the complexity of the surface layers of these stars, including the presence of magnetic fields and chemical inhomogeneities indicates that a more sophisticated modeling of the pulsations is also desirable when attempting to interpret their power spectra. The present work builds on previous studies of the direct effect of the magnetic field on pulsations in the presence of large-scale magnetic fields. 
A detailed discussion of the impact that the results presented here may have on  the power spectra of selected models of roAp stars will be provided in a future paper.

\subsection{Direct and indirect effects of the magnetic field}
\label{subs:mag}l
The strongest magnetic field so far detected in a cool Ap star has a mean magnetic field modulus of $<B>=25.5$kG \citep{hubrig05}. Generally speaking, however,  roAp stars are permeated by only relatively strong, large scale magnetic fields, with typical magnitudes of a few kG \citep{mathys97,hubrig04}. These magnetic fields influence the oscillations both directly, by generating an additional restoring force that affects the wave dynamics in the surface layers, and indirectly, by interfering, and possibly suppressing, envelope convection. In fact, the suppression of envelope convection, at least in some angular region of the star, seems to be a necessary condition for the driving of the observed high frequency oscillations in otherwise standard stellar models of roAp stars \citep{balmforth01,cunha02,saio05}.

Magnetic perturbations to the eigenfrequencies and eigenfunctions of the oscillations in stellar models permeated by an intense, large scale magnetic field have been computed by \citet{dziembowski96}, \citet{bigot00},  \citet{cunha00} and  \citet{saio04}. These authors used a singular perturbation approach, that takes into account the fact that in the surface layers the magnetic effect on the oscillations cannot be treated as a small perturbation \citep{biront82}.  These works revealed some important consequences of the direct effect of the magnetic field on pulsations. In particular, it was found that the eigenfunctions are perturbed to an extent that at the surface they can no longer be described by a single spherical harmonic function. When the cancellation effects originated by our incapacity to resolve the star are taken into account, it is clear that this can lead to an erroneous identification of the degree of the mode that is observed. In fact, if excited, it is possible that modes of degree significantly higher than $l=3$ may generate lower degree components near the surface that may be observed. 

Another important result was the finding that for certain magnetic field intensities and oscillation frequencies the pulsation energy losses are particularly high. These energy losses may contribute to the selection effects that determine the frequency range in which unstable high frequency oscillations may be found, for a given magnetic field. Finally, the maxima of energy losses mentioned above were also found to be associated with abrupt changes in the oscillation frequencies. These abrupt changes may introduce anomalies in the power spectra of roAp stars, as was suggested to happen in one of the best studied roAp stars HR1217 \citep{cunha01}. Such abrupt changes will be discussed in more detail in section \ref{results}.

In  \citet{cunha00} a variational approach to the calculation of the frequency shifts was applied to polytropic models of roAp stars. In the present paper we extend their work, applying the same approach to standard stellar models permeated by magnetic fields of different configurations. As before, only the direct effect of the magnetic field on pulsations is taken into account. In section \ref{m-a} we set the problem and discuss the physics of the pulsations under study in the different regions of the star. In section \ref{model} we describe the equilibrium model and the magnetic field configurations that will be considered.

 In section \ref{vp} we recall the expression derived for the magnetic frequency perturbations under the variational approach used by \cite{cunha00} and in section \ref{results} we discuss the results obtained in particular for a) different boundary conditions applied to the pulsations near the surface and b) different magnetic field configurations. A comparison with the results obtained by Saio (private communication) for the same model is also presented in that section. Finally, in section \ref{conclusions} we summarize our most relevant results.

\section{Magnetoacoustic waves}
\label{m-a}
\subsection{Solutions}
Adiabatic motions in a non-rotating plasma in the limit of perfect conductivity
are described by the system of magnetohydrodynamic equations,
\begin{eqnarray}
\frac{\partial{\bf B}}{\partial t}=\nabla\wedge\left({\rm{\bf v}}\wedge
{\bf B}\right),
\label{eq:induction}
\end{eqnarray}
\begin{eqnarray}
\frac{D\rho}{Dt}+\rho\nabla\cdot{\rm{\bf v}}=0,
\end{eqnarray}
\begin{eqnarray}
\rho\frac{D{\rm{\bf v}}}{Dt}=-\nabla p+{\bf j}\wedge{\bf B}+\rho{\bf g},
\label{eq:momentum0}
\end{eqnarray}
\begin{eqnarray}
\frac{Dp}{Dt}=\frac{\gamma p}{\rho}\frac{D\rho}{Dt},
\end{eqnarray}
\begin{eqnarray}
\nabla\cdot{\bf B}=0,
\label{eq:j}
\end{eqnarray}
where the current density ${\bf j}$ is given by
\begin{eqnarray}
{\bf j}=\frac{1}{\mu_0}\nabla\wedge{\bf B},
\label{eq:j}
\end{eqnarray}
and $\mu_0$ is the permeability of the vacuum, $\rho$ is the density, $p$ is the pressure, ${\bf g}$ is 
the gravitational field, ${\bf B}$ is the vector magnetic field, ${\rm{\bf v}}=\frac{{\partial}{\mbox{\boldmath $\xi$}}}{\partial t}$ is the velocity, ${\mbox{\boldmath $\xi$}}$ is the vector displacement, $\gamma$ is the first adiabatic exponent and $t$ is the time.

In a star permeated by a magnetic field, the direct contribution of the field to the dynamics of what would otherwise be acoustic pulsations is important only near the surface, where the magnetic pressure is comparable, or larger, than the gas pressure. We shall refer to that region as {\it magnetic boundary layer}. 
%In the interior the dynamics is effectively field-free and the acoustic oscillations are described by the usual equations of hydrodynamics. Thus, to study the magetic perturbations to the acoustic oscillations the magnetohydrodynamic equations need to be solved only in the magnetic boundary layer. The solutions at the base of the magnetic boundary layer, which is set in a region where the gas pressure doninates, may then be used to define boundary conditions to the equations describing the acoustic oscillations in the interior.

The system of equations (\ref{eq:induction})-(\ref{eq:j}) is solved, for the magnetic boundary layer, in  a local, plane-parallel approximation, assuming small perturbations to the equilibrium structure, which is assumed to be permeated by a force free magnetic field.  The (Eulerian) perturbation to the gravitationl potential is ignored (Cowling approximation). Moreover, only high order, low degree, axisymmetric modes will be considered. The equations that are actually solved are described in appendix A. In the present section we attempt only to give some insight into particular classes of solutions that might be found  under the conditions that are of interest to us.  

Let us consider first the simple case of a homogeneous fluid, permeated by a uniform magnetic field, in conditions such that the magnetic and gas pressure are comparable. Under these conditions, two types of linear, adiabatic, non-diffusive magnetoacoustic perturbations may be identified, namely, the fast and the slow magnetoacoustic waves \citep[e.g.][]{priest82}. These waves correspond to two distinct solutions of the equations that govern the perturbations in such a homogeneous fluid. They may be regarded as an acoustic wave modified by the magnetic field and as a compressional Alfv\'en wave modified by the gas pressure. 

When the density is stratified, as in a real star, it is generally no longer possible to identify the two distinct solutions described above. Instead, the perturbed equations describe a single, rather more complex, magnetoacoustic wave. However, in the outermost layers of the star, where the gas pressure is very small compared to the magnetic pressure, and below the surface layers, where the gas pressure dominates over the magnetic pressure, the magnetoacoustic wave decouples into waves that are essentially magnetic and waves that are essentially acoustic. Thus, while the equations describing the linear, non-diffusive, magnetoacoustic perturbations have to be solved numerically throughout the region where the magnetic and gas pressure are comparable, approximate analytic solutions can be found to describe the decoupled acoustic and magnetic waves both in the upper atmosphere and in the interior of the star. 

\subsection{Boundary conditions}
Some of the boundary conditions applied in the present work to solve the perturbed equations in the magnetic boundary layer make use of the decoupling described above. In particular the numerical solutions are matched into approximate analytical solutions both in the outer layers of the star and in the interior. The details of these and of the remaining boundary conditions applied are given in appendix B. Nevertheless, it is worth noting here that two possibilities are considered regarding the reflection/transmission of the waves in the surface layers. In some cases a mechanical boundary condition is applied which assures that the waves are fully reflected at the outermost layer of the model, regardless of the oscillation frequency. In other cases, the component of the displacement parallel to the magnetic field direction in the atmosphere is matched onto an outgoing propagating wave whenever the oscillation frequency is above the acoustic cutoff frequency. This is justified by the fact that in the upper atmosphere the component of the solution parallel to the magnetic field direction is essentially an acoustic wave (Sousa $\&$ Cunha - private communication) and, thus, its energy is expected to be lost when the oscillation frequency is above the acoustic cutoff frequency. We recall that the acoustic cutoff frequency appropriate for the oscillations under study depends on the inclination of the magnetic field, being equal to the cut-off frequency in the absence of the magnetic field when the direction of the latter is locally vertical and tending to zero when the direction of the local magnetic field tends to horizontal \citep{dziembowski96,bigot00}.

\section{Models}
\label{model}
The equilibrium model used in this work was computed by Marques et al.\footnote{models available at {\it www.astro.up.pt/corot/models/cesam/}} with the evolutionary code CESAM \citep{morel97} . The atmosphere of the model was extended to a minimum density of $10^{-11} gcm^{-3}$.  The global parameters of the model are given in table~\ref{tab:model}. 
%%%%%%%%%%%%%%%%%%%%%%%%%%%%%%%%%%%%%%%
\begin{table}
\begin{center}
   \caption{global parameters of the model used in the computations. 
$M/{\rm M}_\odot$ and $R/{\rm R}_\odot$ are the mass and radius in solar units, $T_{\rm eff}$ is the effective temperature and $\log(L/{\rm L}_\odot)$ is the logarithm to the base 10 of the luminosity in solar units.}
    \vglue 20pt
    \begin{tabular}[h]{ccccc}
\hline
 Model & $M/{\rm M}_\odot$ & $R/{\rm R}_\odot$ & $T_{\rm eff}$ & $\log(L/{\rm L}_\odot)$ \\
\hline  
Cesam - {\small ZAMS} & 1.8 & 1.57 &  8362 K & 1.035\\
\hline
   \end{tabular}
   \label{tab:model}
\end{center}
\end{table}
%%%%%%%%%%%%%%%%%%%%%%%%%%%%%%%%%%%%%%

Since the magnetic field is assumed to be force free, it does not influence the stratification of the equilibrium model. However, when small perturbations to the equilibrium structure are considered, the perturbed magnetic field responds through the Lorentz force, ${\bf j}\wedge{\bf B}$, present in equation (\ref{eq:momentum0}). Thus, we shall refer to 'the magnetic model' and 'the non-magnetic model' whenever we compare the oscillations in the presence of a magnetic field with the oscillations in an otherwise similar model without a magnetic field. Moreover, the terms perturbed and unperturbed will hereafter be used when referring to the properties of the oscillations in the magnetic and non-magnetic models, respectively.

 The influence that the magnetic response has on what would otherwise be an acoustic oscillation, depends, among other factors, on the configuration of the magnetic field. 
Three different magnetic field configurations are considered in the present work: a dipolar and a quadrupolar magnetic field defined, respectively, by
\begin{eqnarray}
\smash[b]{\bf B_{\rm d}}=b_{\rm d}/r^3(\cos\theta\,\, {\rm{\bf e_r}}+1/2\sin\theta\,\, {\rm{\bf e_\theta})}, \nonumber
\end{eqnarray}
 and
\begin{eqnarray}
\smash[b]{\bf B_{\rm q}}=b_{\rm q}/r^4(1/2\,(3\cos^2\theta-1)\,\, \rm{\bf e_r}+\cos\theta\sin\theta\,\, {\rm{\bf e_\theta)}},\nonumber
\end{eqnarray}
and a magnetic field, ${\bf B_{\rm dq}}$, composed of aligned dipolar plus quadrupolar components. Here  $b_{\rm d}$ and $b_{\rm q}$ are constant amplitudes, $r$ and $\theta$ are the usual coordinates in the spherical coordinate system $(r,\theta,\phi)$ and $\rm{\bf e_{\rm r}}$ and $\rm{\bf e_\theta}$ are the corresponding unit vectors. 

Since the magnetic perturbations involve only quadratic terms in the magnetic field, the perturbations induced on the eigenfunctions by dipolar or quadrupolar magnetic fields are necessarily symmetric about the magnetic equator. Thus, when expanded in spherical harmonics, the perturbed eigenfunctions will include only degrees $l$ of parity identical do the degree of the unperturbed mode. On the other hand, the effect of a magnetic field whose configuration is the sum of dipolar and quadrupolar components does not comply with the referred symmetry properties. Consequently, unlike the dipolar and quadrupolar cases, the third magnetic field considered here can perturb the displacement eigenfunctions in a way such that their expansion in spherical harmonics contains degree components of different parity.

\section{Magnetic perturbations: variational approach}
\label{vp}
The direct effect produced by the magnetic field on the wave dynamics in the outer layers of the star modifies the frequencies of the oscillations. We define the magnetic frequency perturbations as the difference between the eigenfrequencies appropriate to the magnetic and the non-magnetic models. Different approaches to the calculation of these frequency perturbations are possible. Here we follow the variational approach used in \cite{cunha00}. 

In the interior of the magnetic model the magnetic field has a negligible contribution to the restoring force associated with the fast wave. There, the dynamics is effectively field-free and the acoustic oscillations are described by the usual equations of hydrodynamics. Consequently, when compared with the acoustic wave propagating in the interior of the non-magnetic model, the fast wave in the interior of the magnetic model differs only due to the effects introduced by the overlying magnetic boundary layer. 

 In the asymptotic limit valid for modes of high radial order, the radial component of the displacement in the non-magnetic model is given, in the region of propagation, by \citep[e.g.][]{gough93}
\begin{eqnarray}
\xi_r\left(r,\theta,\phi,t\right)\sim A\frac{\kappa^{1/2}}{r\rho_0^{1/2}}\cos\left(\int_{ r}^{ R^*} \kappa {\rm d}r+\delta\right)Y_l^m {\rm e}^{{\rm i}\omega t},        
\label{eq:xi} 
\end{eqnarray} 
where $\rho_0$ is the density in the equilibrium model, $\kappa$ is the vertical acoustic wavenumber, $\omega$ is the oscillation frequency, $A$ is a 
constant, $\delta$ is a phase, $R^*$ is a particular value of the radial coordinate and $Y_l^m$ is a spherical harmonic of degree $l$ and order $m$. 

In the magnetic model, the spherical symmetry of the problem is broken by the effect produced by the magnetic field on the oscillations in the magnetic boundary layer. Consequently, the dependence on latitude of the radial displacement in the interior of the magnetic model is no longer given by a single spherical harmonic of degree $l$. However, the fact that the magnetic field varies only on large scales and that the modes to be considered are of low degree allows the problem to be solved locally, assuming at each latitude a locally uniform field (see appendix A for details). Under this assumption the radial component of the displacement below the magnetic boundary layer (but still sufficiently close to the surface), may still be expressed, at each particular latitude, by equation~(\ref{eq:xi}). 

The effect of the magnetic boundary layer on the oscillations in the interior may be seen as a shift in the phase $\delta$ computed at the base of that layer (defined to be $r=R^*$), which changes from the value $\delta_{\rm unp}$ it would have in the non-magnetic model to a new value $\delta$ which depends on latitude. Moreover, the vertical acoustic wavenumber and the oscillation frequency are also modified.

In the variational approach the phase shifts $\Delta\delta\left(\theta\right)=\delta-\delta_{\rm unp}$ are used to determine the magnetic perturbations to the eigenfrequencies. In practice the variational principle is applied to determine the perturbations to the eigenfrequencies without having to determine the perturbations to the eigenfunctions. According to \cite{cunha00} the difference between the angular frequencies in the magnetic and non-magnetic models, for a mode of degree $l$ and azimuthal order $m$, is given by:
     
\begin{eqnarray}
\frac{\Delta\omega}{\omega}\simeq-\frac{\overline{\Delta\delta}}
{{\omega}^2\int_{r_1}^{R^*}{c_0}^{-2}\kappa^{-1}{\rm d}r}   
\label{eq:fshift},
\end{eqnarray}
%****************************************************************************
where $c_0$ is the sound speed in the equilibrium model, $r_1$ is the lower turning point (at which $\kappa=0$)
and $\overline{\Delta\delta}$ is the integral phase shift, i.e.
\begin{eqnarray}
\overline{\Delta\delta}=\frac{\int_{0}^{\pi}\Delta\delta\left({Y_l}^m\right)^2\sin
\theta{\rm d}\theta}{\int_{0}^{\pi}\left({Y_l}^m\right)^2\sin\theta{\rm d}\theta} 
\label{eq:idd}.
\end{eqnarray}
%-------------------------------------
From this expression we can calculate the shift in the corresponding cyclic frequency through the relation $\Delta\nu= (2\pi)^{-1}\Delta\omega$. The base of the magnetic boundary layer, $R^*$, (at which the solutions appropriate to the magnetic boundary layer are matched onto those appropriate to the interior) must be placed in the region where the gas pressure dominates over the magnetic pressure, but sufficiently close to the surface for the plane-parallel approximation to be valid. In the present calculations we have used $R^* / R = 0.98$

At certain frequencies the eigenfunctions might be significantly perturbed by the magnetic field, at least in some angular region of the star (cf.\ works mentioned in subsection \ref{subs:mag}). This fact raises some concern regarding the accuracy of the results obtained for those frequencies, independently of which approach (variational or expansion in spherical harmonics) is used.

If the eigenfunctions are strongly perturbed in a relatively small range of latitudes, the spherical harmonic expansion used in the works of \citet{dziembowski96}, \citet{bigot00} and \citet{saio04} may be 'contaminated' by rather high $l$ terms. These terms might be completely missed if a simple convergence criteria is used. Such problems were acknowledge in \citet{saio04}, who reported convergence problems at some values of the unperturbed frequency.  

In the case of the variational approach, the unperturbed eigenfunctions in the interior are used to calculate the perturbations to the eigenfrequencies.
As argued by \citet{cunha00}, even though the eigenfunctions might be strongly modified in the outer layers, in the interior, where most of the inertia of the modes resides, the perturbation to the eigenfunctions is likely to be sufficiently small to justify the use of a first order variational approach. However, the possibility that the sharp variations found by \citet{cunha00} in the phase shifts might trap the modes in particular angular regions of the star (Montgomery $\&$ Gough - private communication) raises some worries. If this trapping takes place  the eigenfunctions in the interior of the magnetic and non-magnetic models will be significantly different. With this in mind Montgomery $\&$ Gough are investigating on the possibility of determining the magnetically perturbed eigenfunctions without recourse to a spherical harmonic expansion. In a future work we plan to use their results to test the accuracy of the frequency shifts derived with the variational approach applied here as well as to determine the corresponding higher order corrections to the latter.

\section{Results and analysis}
\label{results}

\subsection{Effect of boundary condition at the top}

\begin{figure*}
\begin{center}
\includegraphics[width=135mm]{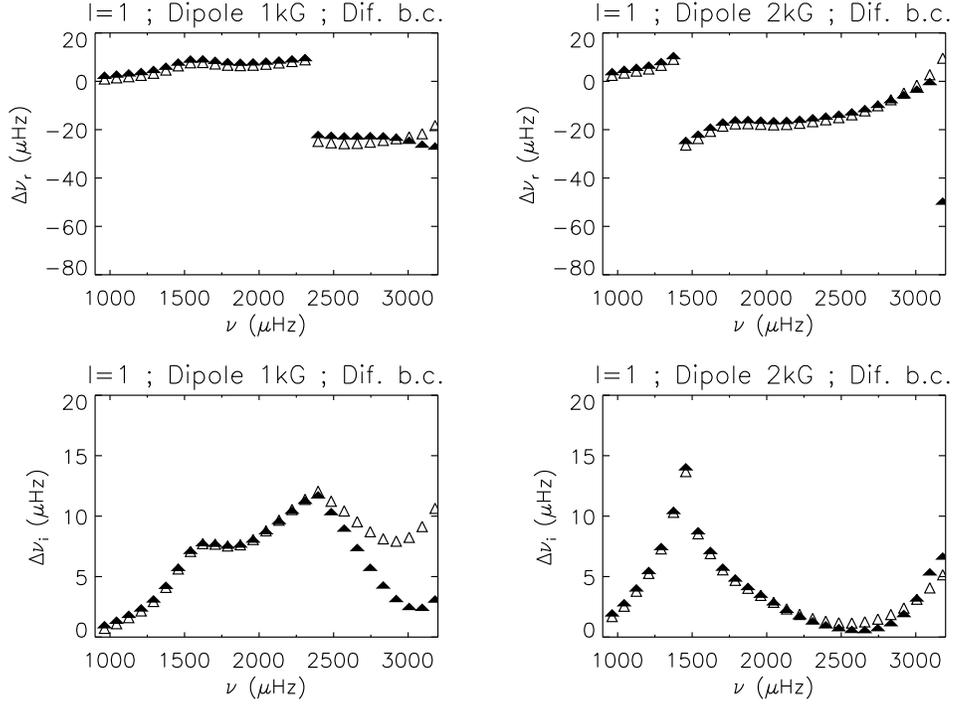}
\caption{Comparison of the real (top) and imaginary (bottom) parts of the frequency shifts, as function of the cyclic frequency $\nu$,  when two different boundary conditions at the outermost layer are applied: {\it full reflection} (filled symbols) and {\it partial transmission}~(open symbols) at the top. Results are shown for two magnetic field polar intensities, namely $B_p=b_d=1$kG (left) and $B_p=b_d=2$kG (right).} 
\label{fig0}
\end{center}
\end{figure*}

\begin{figure*}
\begin{center}
\includegraphics[width=135mm]{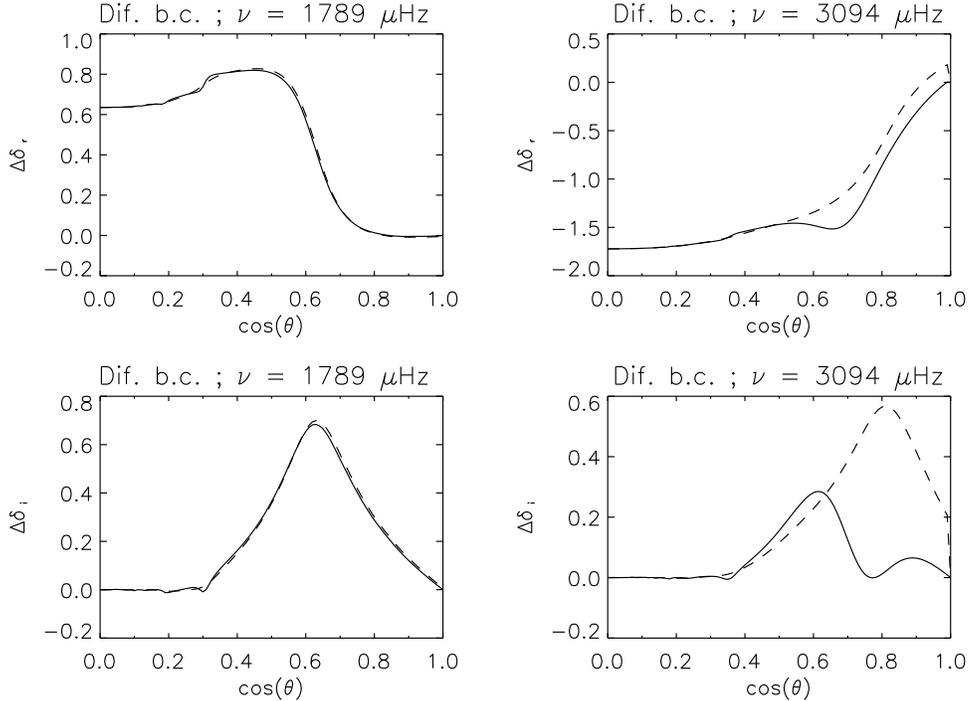}
\caption{Comparison of the real (top) and imaginary (bottom) parts of the phase shifts when two different boundary conditions at the outermost layer are applied for two different oscillation frequencies.  The continuous line shows the case of {\it full reflection} and the dashed line shows the case of {\it partial transmission} at the top.  Results are shown for a dipolar magnetic field with polar intensity $B_p=b_d=1$KG, for two different unperturbed frequencies, namely, $\nu=1789 \;\mu$Hz (left) and $\nu=3094 \;\mu$Hz (right). The cutoff frequency for acoustic modes in the absence of a magnetic field is $\nu_c \approx 2550 \;\mu$Hz for the model used.}
\label{fig1}
\end{center}
\end{figure*}

Figures \ref{fig0} and \ref{fig1} compare, respectively, the frequency and the phase shifts that are obtained with a dipolar magnetic field, when two different boundary conditions are applied to the acoustic component of the wave at the outermost layer of the star.  For the phase shifts only the range of co-latitudes $0 < \theta < \pi / 2$ is shown since for magnetic fields with a dipolar configuration the perturbations are symmetric about the magnetic equator.

In figure \ref{fig0} it is seen that the boundary condition applied might be particularly important for the determination of the imaginary part of the oscillation frequency, if the frequency of the mode is above the acoustic cutoff frequency appropriate to the non-magnetic model, which in the present case is $\nu_c\approx 2550 \mu$Hz. Comparing the results for two different magnetic field intensities (left and right hand side panels of figure  \ref{fig0}) we also see that the boundary condition applied has a larger impact on the results when the imaginary part of the frequency shifts is close to a maximum than when it is close to zero.

For the phase shifts, two oscillation frequencies are shown, namely, below (left panels) and above (right panels) $\nu_c$.
When the oscillation frequency is below $\nu_c$ the phase shifts are almost independent of the boundary condition applied. This was to be expected near the magnetic poles. There, the magnetic field is nearly vertical, and  the critical frequency for the acoustic component of the wave (that may be identified in the outer magnetically dominated layer) is similar to $\nu_c$. Thus, for frequencies below $\nu_c$ this component of the wave is reflected independently of which boundary condition is applied and, as a result, the phase shifts obtained are very similar. Away from the magnetic poles, however, the critical frequency for the acoustic component of the wave decreases \cite[e.g.][]{bigot00}, tending to zero as the magnetic field direction becomes closer to horizontal. The fact that the phase shifts are very similar also for higher co-latitudes indicates that the acoustic component of the wave propagating away in the outer layers has a relatively small effect on the acoustic oscillation in the interior when the magnetic field is significantly inclined. 

When the oscillation frequency is above $\nu_c$ a significant difference is found between the phase shifts obtained  with the two different boundary conditions. This difference is significant up to co-latitudes of about 60$^0$, becoming negligible as the magnetic field direction becomes closer to horizontal.

\subsection{Different magnetic field configurations}

\begin{figure*}
\begin{center}
\includegraphics[width=125mm]{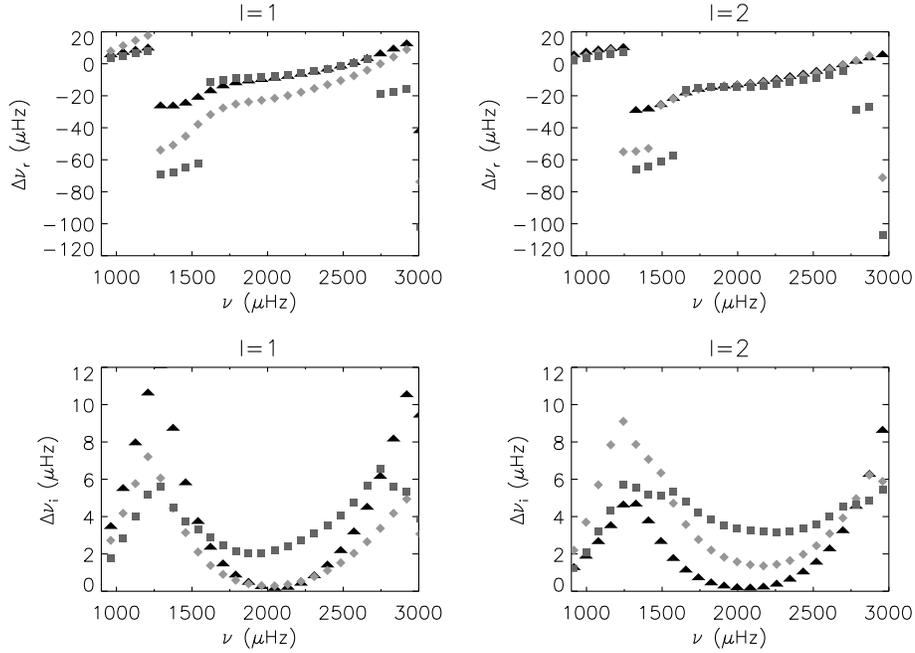}
\caption{Real (top) and imaginary (bottom) parts of the frequency shifts, as function of unperturbed frequency, for magnetic fields with polar strength of $B_p~=~3$~kG and different configurations - {\it dipole} (triangles), \, {\it quadrupole} (diamonds), \, {\it aligned dipole $+$ quadrupole} (squares). The results are shown for modes of degrees $l=1$ (left) and $l=2$ (right). In the case of a magnetic field composed of a dipolar plus a dipolar component both components have the same polar strength, which is such as to make the total polar strength of the magnetic field is $B_p~=~3$~kG.
}
\label{fig3}
\end{center}
\end{figure*}

\begin{figure*}
\begin{center}
\includegraphics[width=125mm]{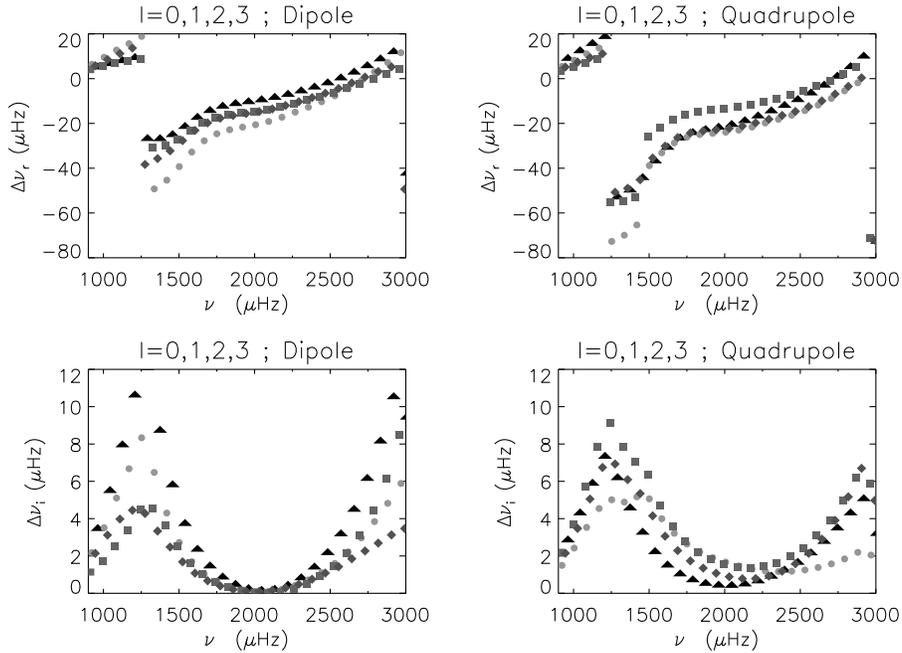}
\caption{Real (top) and imaginary (bottom) parts of the frequency shifts, as function of unperturbed frequency, for modes of different degree - $l=0$ (circles), \, $l=1$~(triangles), \,$l=2$~(squares), \,$l=3$~(diamonds). Results are shown for magnetic fields with dipolar (left) and  quadrupolar (right) configuration and a polar strength of $B_p~=~3$~kG.} 
\label{fig4}
\end{center}
\end{figure*}

Figures \ref{fig3} and \ref{fig4} show, respectively, the frequency shifts for different magnetic field configurations, all with the same polar strength, and the frequency shifts for modes of different degree. The frequency shifts follow a similar trend in all cases: as the radial order increases, the frequency perturbations generally increase smoothly. At some particular frequencies, however, the real part of the frequency perturbation jumps by an amount that is significantly larger than the mode-to-mode variations seen in the smooth trend. At the same unperturbed frequencies, the imaginary part of the perturbation reaches a maximum.  This behavior is similar to that first found in this context by \citet{cunha00}, for polytropic models, and later confirmed for more sophisticated stellar models by \citet{saio04}. 

There are some additional facts that become evident in the present results, namely,
\begin{itemize}
\item the jumps in the frequency shifts may be positive or negative, i.e., they may increase or decrease the frequency of the modes in comparison to what would be expected if the smooth trend were to be followed;
\item the frequencies at which the first maxima of energy losses take place {\it do not depend} on the degree of the modes nor on the magnetic field configuration (at least for the magnetic field configurations considered here);
\item the amount by which the real part of the frequency shifts change at the jumps  {\it depends} both on the magnetic field configuration and on the degree of the modes.
\end{itemize}

\begin{figure}
\begin{center}
\includegraphics[width=84mm]{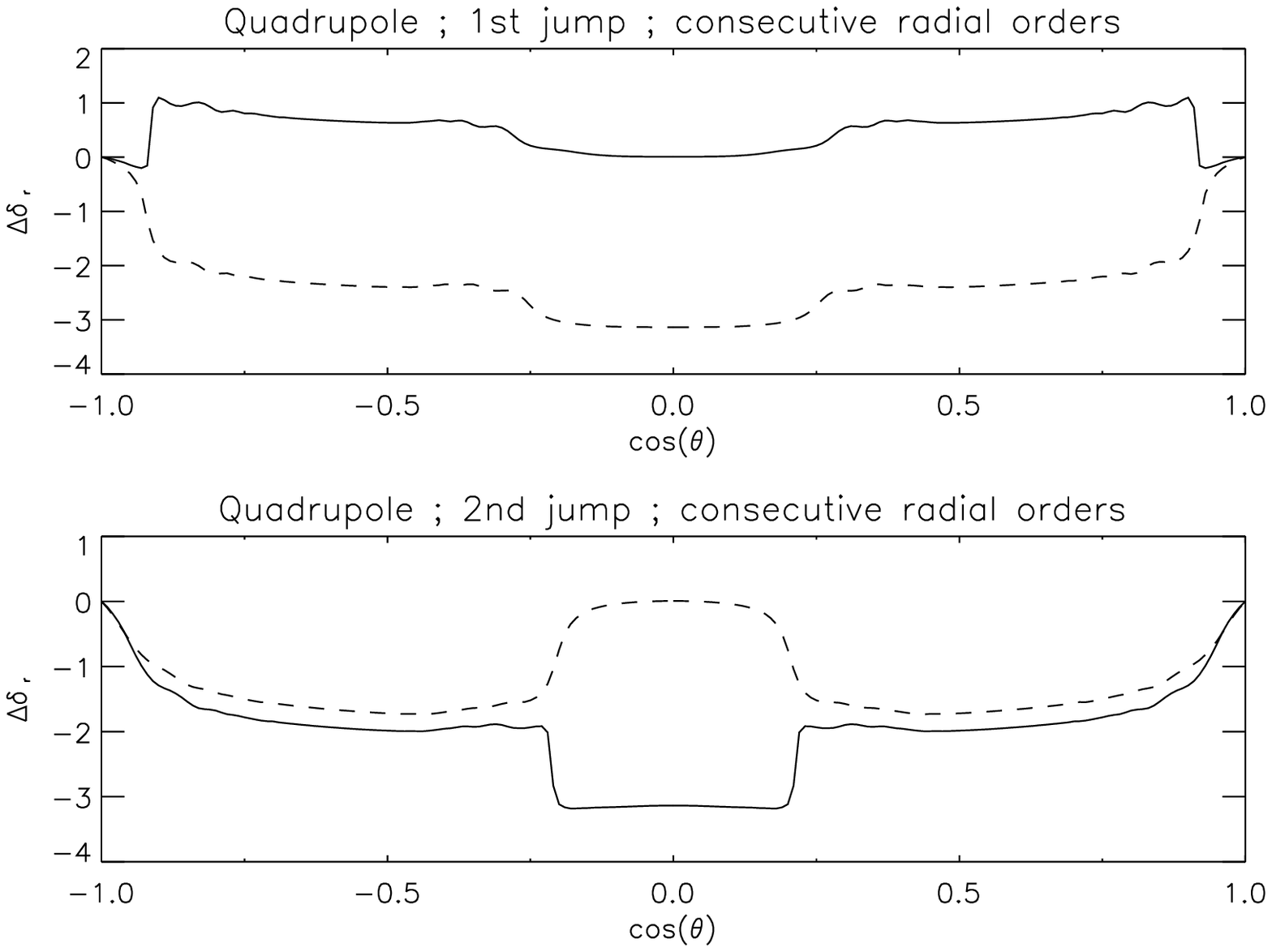}
\caption{Real part of the phase shifts as function of co-latitude $\theta$ for two modes of consecutive radial order (and degree $l=1$) in a model permeated by a quadrupolar magnetic field of polar strength $b_q=3$kG. The upper panel shows the results for the modes with frequencies immediately below (continuous line) and immediately above (dashed line) the first frequency jump while the lower panel shows the results for modes with frequencies immediately below (continuous line) and immediately above (dashed line) the second frequency jump. 
} 
\label{fig5}
\end{center}
\end{figure}

\begin{figure}
\begin{center}
\includegraphics[width=84mm]{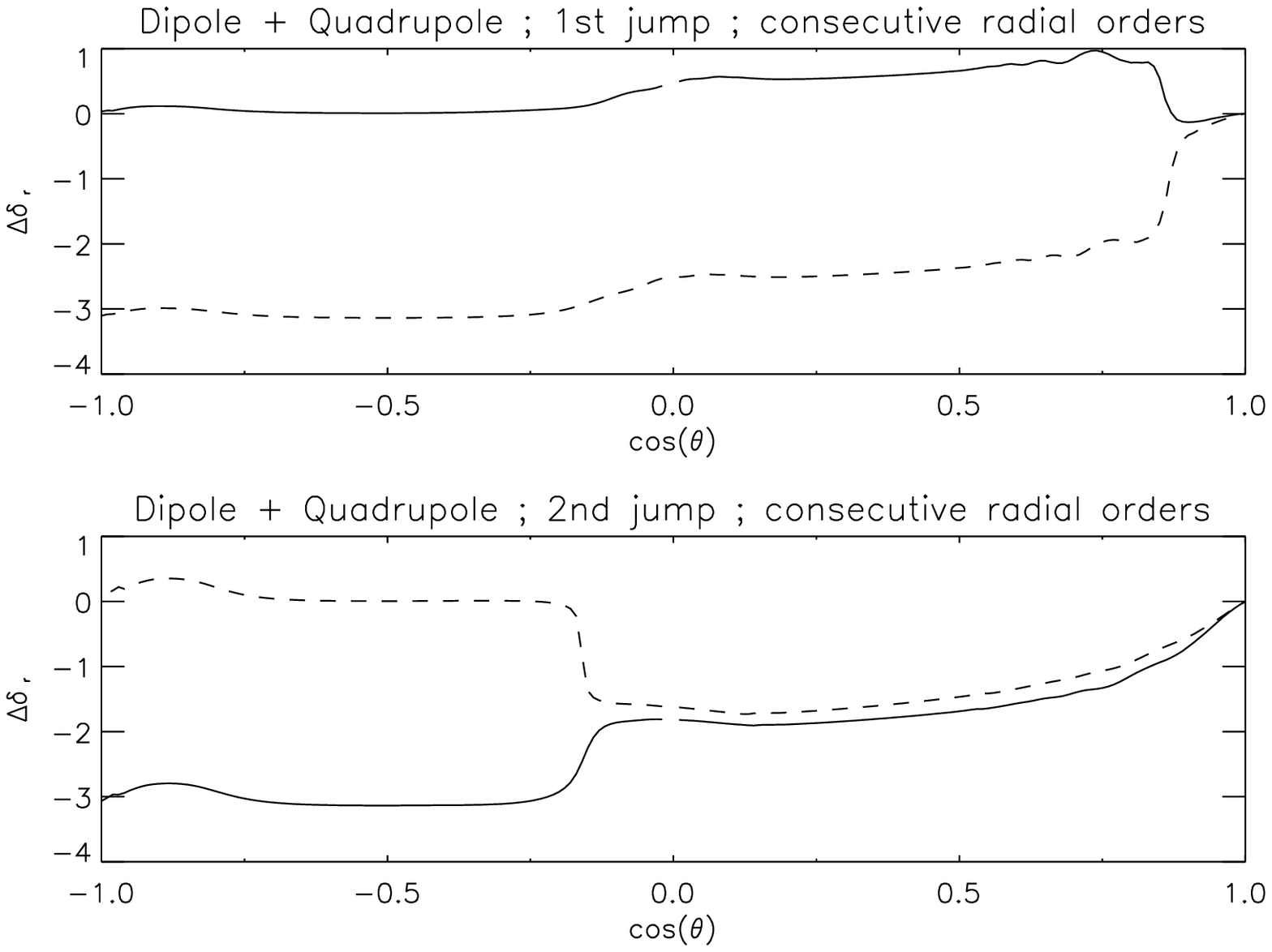}
\caption{Same as figure \ref{fig5} but for a magnetic field which configuration is the sum of aligned dipolar and quadrupolar components with the same polar strengths.  
} 
\label{fig6}
\end{center}
\end{figure}

Figures \ref{fig5} and \ref{fig6} show the real part of the phase shifts obtained for two modes of consecutive radial order in models permeated, respectively, by a quadrupolar magnetic field and a magnetic field composed of aligned dipolar and quadrupolar components.

Each figure has two panels and in each of the panels the two modes shown are chosen such that their frequencies are, respectively, below and above a frequency jump. It is readily seen from these figure that the sudden changes in the real part of the frequency perturbations result from jumps of $\approx\pi$ in the real part of the phase shifts, which occur at some well defined co-latitudes. \citet{cunha00} noted that for polytropic models with a dipolar magnetic field the co-latitude at which the real part of the phase shifts jumped was approximately the co-latitude at which the product $B_xB_z$ was maximal. For magnetic fields with dipolar configuration that co-latitude is $\theta=45^0$ and corresponds to the co-latitude at which $|B_x|=1/2 |B_z|$.

Figures \ref{fig5} and \ref{fig6} show that the latter relation holds also for magnetic fields with different configurations, in more sophisticated stellar models. Because a quadrupolar magnetic field satisfies the condition $|B_x|=1/2 |B_z|$ twice in the range of co-latitudes $0 <\theta < \pi / 2$, two different jumps in the real part of the phase shifts are found, one at $\theta\approx 24^0$ and one at $\theta\approx 78^0$, at different oscillation frequencies. These phase jumps are responsible for the two frequency jumps that may be seen in the frequency range $1000\mu$Hz~$< \nu < 2000 \mu$Hz in the upper right hand side (rhs) panel of figure \ref{fig4}. The fact that the second frequency jump is apparent only for modes of even degrees (\ $l=0$ (circles) and $\l=2$ (squares)) is a direct consequence of the high value of $\theta$ at which the real part of the phase shift jumps in this case. Modes of odd degree hardly feel the second jump in the phase shift because their amplitudes are very low at the co-latitudes at which the phase shifts of the two consecutive modes is significantly different.

 In the case of a magnetic field composed of aligned dipolar plus quadrupolar components the phase shifts are not symmetric about the magnetic equator. Hence, unlike the case of dipolar and quadrupolar magnetic fields, for this configuration jumps in the phase shifts may take place in each hemisphere separately. In figure  \ref{fig6} it is possible to see the first two jumps in the phase shifts for the case of a magnetic field composed of aligned dipolar and quadrupolar components of the same polar strength. The first jump in the phase shift takes place at $\theta\approx 31.9^0$ while the second takes place at $\theta\approx 100.3^0$. In both cases the condition $|B_x|= 1/2 |B_z|$ is satisfied. These jumps in the phase shifts are responsible for the two frequency jumps seen in the squared symbols of the upper panels of figure \ref{fig3}, in the frequency range $1000\mu$Hz~$< \nu < 2000 \mu$Hz.

\subsection{Cyclic behaviour and scaling}

\citet{campbell86} noted that the equations describing the wave dynamics in the magnetic boundary layer of a polytropic model are invariant under a particular scaling of the variables involved. As argued in \citet{cunha00}, this scaling implies that in polytropic models the frequency shifts obey the relation $\Delta\nu\approx f(B_p^{1/4}\nu)$, where $f$ is some function and $B_p$ is the polar strength of the magnetic field. In practice, this means that the frequency perturbations scale in a very particular way with the magnetic field and the oscillation frequency.

\citet{cunha00} also found that in polytropic models the magnetic perturbations vary cyclically with the oscillation frequency. Given the scaling mentioned above, a cyclic behaviour with the magnetic field intensity should also be expected. In their stellar models \citet{saio04} found both a scaling with magnetic field intensity and oscillation frequency, and a cyclic behaviour of the frequency shifts. The scaling the authors found for the frequency shifts in their models, namely, $f(B_p^{0.7}\nu)$, differs, in the exponent, from the scaling found in polytropes. As explained in their paper, this difference reflects the difference in the structure of the outer layers of the two models.

\begin{figure*}
\begin{center}
\includegraphics[width=125mm]{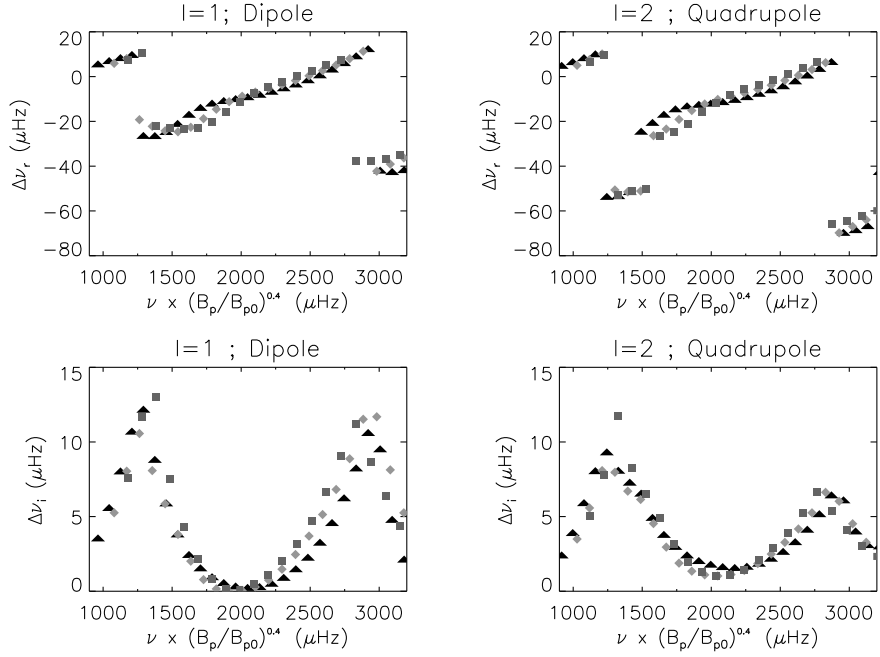}
\caption{Scaling of the real (top) and imaginary (bottom) parts of the frequency shifts with magnetic strength, according to the relation $\nu(B_p/B_{p0})^{0.4}$, where $B_p = 3$~kG~(triangles), \, $B_p = 4$~kG~(diamonds), \, $B_p = 5$~kG~(squares).  In this figure we have used \,\,$B_{p0}~=~3$~kG.}
\label{fig7}
\end{center}
\end{figure*}

In the present models the frequency shifts also display an approximate cyclic behaviour, which may be seen in figure \ref{fig7}. A scaling of the frequency perturbations with magnetic field and oscillation frequency, of the type $f(B_p^{0.4}\nu)$, is also evident in the same figure.  The exponent of the scaling is different from both the exponent found in the polytropic models and in \citet{saio04}, which was to be expected given that the outer layers of the model used here are different from the outer layers of both those models. 

We should note that the scaling shown in figure \ref{fig7} seems to be valid only in the range of magnetic field intensities shown, namely, 3-5 kG. To adjust the frequency shifts obtained with less intense magnetic fields to the curve shown in figure \ref{fig7} we would need to change the exponent in the scaling again. This is not surprising, however, since as argued by \citet{saio04} the exponent in the scaling reflects the average structure of the magnetic boundary layer. The extent of the magnetic boundary layer depends on the magnetic field strength (reaching deeper layers if the magnetic field is more intense). Thus, it is not surprising that the scaling with the magnetic field (with fixed exponent) breaks down when the range of magnetic field intensities considered gets too wide.

\subsection{Comparison with Saio's results}
\begin{figure*}
\begin{center}
\includegraphics[width=125mm]{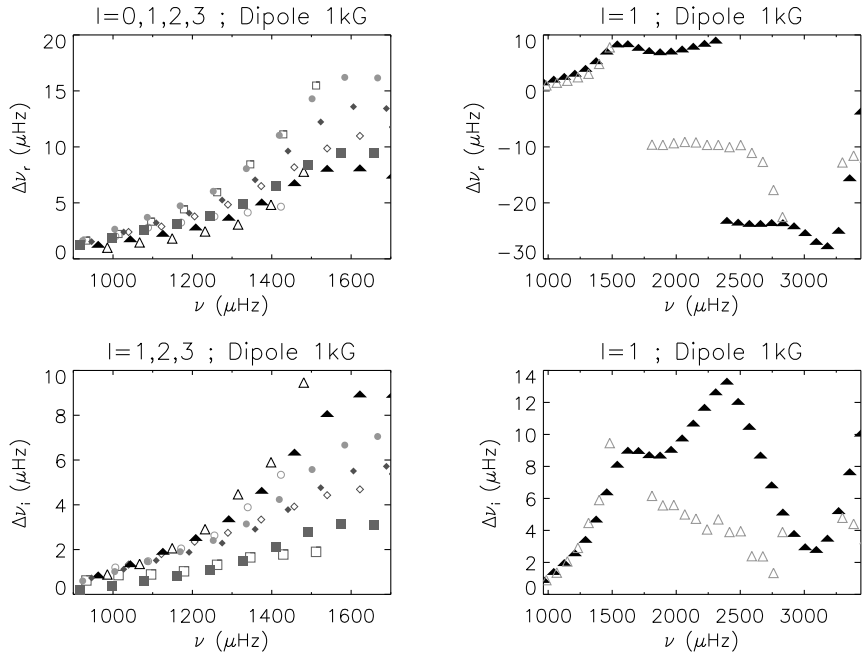}
\caption{Comparison between the real (top) and Imaginary (bottom) parts of the frequency shifts calculated with {\it Saio's code} (open symbols) and {\it Cunha's code} (filled symbols), for the stellar model described in table~\ref{tab:model} and the same magnetic field, with dipolar configuration and polar strength of $B_p=1$kG. The left panels show the comparison for modes of different degrees $l=0$ (circles), \, $l=1$~(triangles), \,$l=2$~(squares), \,$l=3$~(diamonds) before the first frequency jump. The right panels show the compassion for modes of degree $l=1$ in a frequency range including the first frequency jump. }
\label{fig8}
\end{center}
\end{figure*}

To compare our results with those obtained with Saio's code it is important to eliminate the effects brought about by the differences in the models used.
Since the outer layers of the model used by \citet{saio04} were significantly
different from those of the model used in this work, Saio (private communication) has computed new frequency shifts using the model described in table \ref{tab:model} with a dipolar magnetic field. A comparison between the results obtained with the two codes is shown in figure \ref{fig8}. 
 
Although the results computed with Saio's code for the model used here are qualitatively in agreement with our results, there are significant differences in the details. On the lowest left hand side panel we can see that the dependence of the imaginary part of the frequency shifts on the degree of the modes is consistent for both codes (with $l=1$ showing the largest shifts and $l=2$ the smallest). However, the same agreement does not hold for the real part of the frequency shifts. On the right hand side panels we can also see that the frequency shifts computed with the two codes jump at different frequencies. This might result from the differences in the boundary condition applied at the top in the two codes. But probably the most striking difference between the two sets of results is the amount by which the real part of the frequency shifts jump, which is significantly larger in our results. This difference was already stressed in \citet{saio04} for the case of polytropic models. Curiously enough, the frequency shifts seem to come into agreement again when the unperturbed frequency is sufficiently far from the jump.

\section{Conclusions}
\label{conclusions}

In the present paper we investigate on the effect that magnetic fields with different configurations have on the oscillation frequencies of acoustic waves propagating in the interior of roAp stars. We consider both magnetic field configurations which produce perturbations that are symmetric about the magnetic equator (like pure dipolar and pure qudrupolar magnetic fields) and a 'mixed' magnetic field configuration, composed by the sum of aligned dipolar and quadrupolar components, which induces a magnetic perturbation that is not symmetric about the magnetic equator.
The interest of studying magnetic fields that may produce perturbations that are not symmetric about the magnetic equator may be exemplified with the well known roAp star HR 1217. The modes observed in this star are believed to be of alternating even and odd degree. However, when attempting to interpret the multiplet structures seen in the power spectrum of this star one finds that odd degree components need to be included when describing any of the modes as a sum of spherical harmonics. This distortion of the even degree modes into a sum of spherical harmonics that include odd components may not be produced by dipolar or by quadrupolar magnetic fields. However, it  could be produced by a magnetic field which effect on pulsations is not symmetric about the magnetic equator.

The main conclusion of the present paper is that the magnetic frequency perturbations are influenced in two distinct ways, associated with two classes of effects:
\begin{itemize}
\item  the overall pattern followed by the frequency shifts scales with frequency in a way that depends essentially on the structure of the outer layers of the model and on the intensity of the magnetic field. 
\item the amount by which the real part of the frequency shifts jump at well defined frequencies depends essentially on the geometry of the problem, i.e.\ on the magnetic field configuration and on the degree of the mode.
\end{itemize}
We also confirmed that the latter result is a direct consequence of the behaviour of the phase shifts calculated at the bottom of the magnetic boundary layer, which jump by $\pi$ at well defined values of co-latitude, namely the co-latitudes at which $|B_x|=1/2 |B_z|$. This simple dependence of the amount by which the real part of the frequency shifts jump on the geometry of the problem opens an interesting possibility, namely, that of using anomalies that might be found in the power spectra of roAp stars to infer information about the magnetic field configuration and/or the degree of the modes observed.

In the present work we have also investigated on the effect of changing one of the boundary conditions applied at the outermost layer of the stellar model. In particular we have considered the case in which the acoustic component of the wave that may be identified in the magnetically dominated region is allowed to propagate away when its frequency is above the appropriate acoustic cutoff frequency, and the case in which that component of the wave is is fully reflected. 

We found that the boundary condition applied to the acoustic component of the wave in the outer layers of the star influences in a significant way the acoustic wave in the interior only when the oscillation frequency is above the acoustic cut-off frequency of an otherwise similar model with no magnetic field. Moreover, we found that the differences in the results obtained with the two boundary conditions are a consequence of the effect the boundary condition has on the solution when the magnetic field is only moderately inclined in relation to the local vertical coordinate.

\section{Acknowledgments}
I am very grateful to Hideyuki Saio for providing part of the data used in figure \ref{fig8}.
MC is supported by FCT (Portugal) through the grant BPD/8338/2002.
This work was supported by FCT and FEDER (through POCI2010) through theproject POCTI/CTE-AST/57610/2004"

\onecolumn

\begin{appendix}
\section{Boundary-layer equations}

%\subsection*{Boundary-layer equations}

In the magnetic boundary layer we adopt the approximation of a locally uniform (inclined) field, which is essentially force free and which therefore does not influence the stratification of the surface layer. Thus, from equations (\ref{eq:induction})--(\ref{eq:j}), we find that in that layer
low-amplitude adiabatic magnetically nondiffusive pulsations are described 
by the following system of linearized equations: 
\begin{eqnarray}
  \frac{{\partial^2}{\mbox{\boldmath $\xi$}}}{\partial t^2}=-\frac{1}{\rho}\nabla p_1+\frac{\rho_1}{\rho^2}\nabla p+\frac{1}{\mu_0\rho}(\nabla\wedge{\bf B}_1)\wedge {\bf B},   
 \label{eq:momentum}
\end{eqnarray} 
\begin{eqnarray}
p_1=-\mbox{\boldmath $\xi$}{}\cdot\nabla p-\gamma p\nabla\cdot{\mbox{\boldmath $\xi$}},    \label{eq:entropy}
\end{eqnarray} 
\begin{eqnarray}
\rho_1=-{\mbox{\boldmath $\xi$}}\cdot\nabla\rho-\rho\nabla\cdot{\mbox{\boldmath $\xi$}},    \label{eq:mass}
\end{eqnarray} 
\begin{eqnarray}
{\bf B}_1=\nabla\wedge({\mbox{\boldmath $\xi$}}\wedge{\bf B}),       
\label{eq:magflux}
\end{eqnarray} 
where the subscript 1 denotes an Eulerian perturbation and  $\rho$,  $p$, and ${\bf B}$, now stand for the corresponding unperturbed quantities.

We adopt a plane-parallel approximation at each latitude and
longitude of the star.
The local vertical coordinate, {\em z}, is taken to increase outwards, 
and to be zero at the surface of the star. 
The horizontal coordinates, $x$ and $y$, are chosen to form a right-handed
locally Cartesian system with the $z$ axis, orientated such that the $x$ axis 
is parallel to the horizontal component of the magnetic field. Thus we set 
${{\bf B}=\left(B_x,0,B_z\right)}$. Moreover, in accordance with our assumption
of a locally uniform magnetic field, we neglect the derivatives of $B_x$ and 
$B_z$ in equations (\ref{eq:momentum})--(\ref{eq:magflux}).  

We seek solutions of the system of equations 
(\ref{eq:momentum})--(\ref{eq:magflux}) that depend
on {\em x}, {\em y} and time $t$ through the factor 
$\exp{\left[{\rm i}\left(k_xx+k_yy+\omega t\right)\right]}$, where 
${{\bf k}=\left(k_x,k_y,0\right)}$ is the (local) horizontal wave 
number and $\omega$ is the oscillation frequency. ${\bf k}$ can be related to the degree $l$  and the azimuthal order $m$ of the mode considered by means of a standard asymptotic approximation for the spherical harmonics \citep[e.g.,][]{campbell86}. For axisymmetric modes the relation reduces to $k_x \approx L^2 / R$ and $k_y = 0$, where $L=l+1/2$. 

We decompose  the displacement into its vertical component and its horizontal components 
parallel and perpendicular to ${\bf k}$, defined respectively by, 
${\xi_z={\mbox{\boldmath $\xi$}}.{\bf e}_z}$, 
${u={{\mbox{\boldmath $\xi$}}.\bf k}/|{\bf k|}}$ and
${v= {\mbox{\boldmath $\xi$}}.\left({\bf e}_z\wedge{\bf k}\right)/|{\bf k|}}$, 
%****************************************************************************
where ${\bf e}_z$ is a unit vector in the $z$ direction.
%****************************************************************************
Using  equations (\ref{eq:entropy})--(\ref{eq:magflux}) to eliminate  
$p_1,\rho_1$ and ${\bf B}_1$ from  equation (\ref{eq:momentum}), and neglecting
the derivatives of $B_x$ and $B_z$, we obtain
\begin{eqnarray}
-\omega^2\rho u={\rm i}|{\bf k}|W+\frac{1}{\mu_0}{\left({\bf B}\cdot\nabla\right)}^2u-\frac{k_xB_x}{\mu_0 |{\bf k}|}{\bf B}\cdot\nabla\left(\nabla\cdot{\mbox{\boldmath $\xi$}}\right), \label{eq:u}
\end{eqnarray} 
\begin{eqnarray}
-\omega^2\rho v=\frac{1}{\mu_0}{\left({\bf B}\cdot\nabla\right)}^2v+\frac{k_yB_x}{\mu_0 |{\bf k}|}{\bf B}\cdot\nabla\left(\nabla\cdot{\mbox{\boldmath $\xi$}}\right), \label{eq:v}
\end{eqnarray} 
\begin{eqnarray}
-\omega^2\rho \xi_z=\frac{\partial W}{{\partial z}}-g\nabla\cdot\left(\rho{\mbox{\boldmath $\xi$}}\right)-\frac{B_z}{\mu_0}\left[{\bf B}\cdot\nabla\left(\nabla\cdot{\mbox{\boldmath $\xi$}}\right)\right]+\frac{1}{\mu_0}{\left({\bf B}\cdot\nabla\right)}^2\xi_z, 
\label{eq:z}
\end{eqnarray} 
where,
\begin{eqnarray}
W={\mbox{\boldmath $\xi$}}\cdot\nabla p+\left(\gamma p+\frac{{\rm B}^2}{\mu_0}\right)\nabla\cdot{\mbox{\boldmath $\xi$}}-\frac{1}{\mu_0}{\bf B}\cdot\nabla\left({\bf B}\cdot{\mbox{\boldmath $\xi$}}\right)
\label{eq:W}
\end{eqnarray} 
and
\begin{eqnarray}
g=\frac{1}{\rho}\frac{{\rm d}p}{{\rm d}z}. 
\label{eq:g}
\end{eqnarray} 

Next we express the boundary-layer equations in a dimensionless form, by defining
\[
\begin{array}{lllll}
\D{\eta=-\frac{z}{R}}, & \D{\sigma=\frac{\omega}{\sigma_0}}, & \D{{\rm p}=
\frac{p}{p_0}}, & \D{\varrho=\frac{\rho}{\rho_0}}, &
\D{U=\frac{\hat{u}}{R}},  \\
\\
 \D{V=\frac{\hat{v}}{R}}, & \D{Z=\frac{\hat{\xi}_z}
{R}}, & \D{K_x=Rk_x}, &
\D{K_y=Rk_y}, &  
\D{b_i=\frac{B_i}{\left(\mu_0\rho_0{\sigma_0}^2R^2\right)^{1/2}}},  
\label{eq:dvariables}
\end{array}
\]
where $R$ is the stellar radius, $\sigma_0=(GM/R^3)^{1/2}$, $p_0$ and $\rho_0$ are, respectively, the central pressure and density and $\hat{u}$, $\hat{v}$ and $\hat{\xi}_z$ are the z-depend parts of the corresponding components of the displacement defined by  $\left(\hat{u},\hat{v},\hat{\xi}_{z}\right)
e^{{\rm i}\left(k_xx+k_yy+\omega t\right)}=\left(u,
v,\xi_{z}\right)$. 
Introducing the new variables in the system of equations 
(\ref{eq:u})--(\ref{eq:g}), and defining
\[
\mathcal C=\frac{p_0}{\rho_0\sigma_0^2R^2},
\]
we obtain
\[
%\begin{eqnarray}
\left\{
\begin{array}{lll}
b_xb_z\frac{K_x}{K}Z^{\prime\prime}+{\rm i}
\left({\mathcal C}\gamma{\rm p}K+b_x^2\frac{K_y^2}{K}\right)Z^{\prime}
+\left({\rm i}{\mathcal C}{\rm p^{\prime}}K-b_xb_zKK_x\right)Z
+{\rm i}b_xb_zK_yV^{\prime}+b_x^2K_xK_yV \; \:
\\
-b_z^2U^{\prime\prime}+
\left(-\sigma^2\varrho+{\mathcal C}\gamma{\rm p}K^2+b^2K^2-b_x^2K_x^2\right)U
 = 0, & &
\\
\\
-\left({\mathcal C}\gamma{\rm p}+b_x^2\right)Z^{\prime
\prime}
-{\mathcal C}\gamma{\rm p^{\prime}}Z^{\prime}+\left(-\sigma^2\varrho+b_x^2K_x^2+{\mathcal C}\left(\frac{\varrho^{\prime}{\rm p^{\prime}}}{\varrho}-{\rm p^{\prime\prime}}\right)\right)Z
-b_xb_z\frac{K_y}{K}V^{\prime\prime}+{\rm i}b_x^2\frac{K_xK_y}{K}V^{\prime}
+b_xb_z\frac{K_x}{K}U^{\prime\prime} \; \; \; \; \; \; \; \; \; \; \; \; \:
\\
+{\rm i}\left({\mathcal C}\gamma{\rm p}K+
b_x^2\frac{K_y^2}{K}\right)U^{\prime}+\left[{\rm i}{\mathcal C}\left(\left(\gamma-1\right){\rm p^{\prime}}K-\gamma^{\prime}{\rm p}K\right)-b_xb_zKK_x\right]U = 0, & & 
\\
\\
-b_xb_z\frac{K_y}{K}Z^{\prime\prime}
+{\rm i}b_x^2\frac{K_xK_y}{K}Z^{\prime}-b_z^2V^{\prime\prime}+2{\rm i}b_xb_z
K_xV^{\prime}+\left(-\sigma^2\varrho+b_x^2K_x^2\right)V+ 
{\rm i}b_xb_zK_yU^{\prime}+b_x^2K_xK_yU = 0, & &
\end{array}
\right.
%\end{eqnarray}
\]
where prime denotes differentiation with respect to $\eta$ and $K=
\left(K_x^2+K_y^2\right)^{1/2}$. For axisymmetric modes, considered in this work, $K_y=0$ and $K_x=K$. Moreover, in this case the component $V$ of the dimensionless displacement decouples from the components $Z$ and $U$, and the 
 third-order system of second-order complex differential equations given above separates into a second-order system describing the components $Z$ and $U$ and a single second-order equation describing the $V$ component.

The second-order system of second-order complex differential equations describing the components $U$ and $V$ was then transformed into a fourth-order system of first-order complex differential equations, and integrated using a fourth-order-accuracy Runge-Kutta algorithm.
%\end{appendix}

%\begin{appendix}
\section{Boundary conditions}
\subsection*{Surface boundary conditions}

Sufficiently high in the atmosphere, where the magnetic pressure is much larger than the gas pressure, the response to motions that are perpendicular to the magnetic field is dominated by the perturbed Lorentz force. Motions along the magnetic field lines, however, will generate an acoustic response, associated with the gradient of the perturbed gas pressure. Hence, in this region two decoupled waves may be identified: a compressional Alfv\'en wave, which displacement is essentially perpendicular to the direction of the the magnetic field, and an acoustic wave, which displacement is essentially along the latter.

Because the density tends to zero as we move towards the outer layers of the atmosphere, the magnetic field must tend to a vacuum field implying that:
\begin{eqnarray}
(\nabla\wedge{\bf B}_1)_{\rm s} = 0,
%[({\bf B}\cdot\nabla){\mbox{\boldmath $\xi$}}-{\bf B}\nabla\cdot{\mbox{\boldmath $\xi$}}]_{\rm s}=(\nabla\psi_{\rm m})_{\rm s},  
\label{eq:bc1} 
\end{eqnarray}
or equivalently,
\begin{eqnarray}
%(\nabla\wedge{\bf B}_1)_{\rm s} =0
[({\bf B}\cdot\nabla){\mbox{\boldmath $\xi$}}-{\bf B}\nabla\cdot{\mbox{\boldmath $\xi$}}]_{\rm s}=(\nabla\psi_{\rm m})_{\rm s},  
\label{eq:bc1_1} 
\end{eqnarray}
where $\psi_{\rm m}$ is an external magnetic potential, regular at infinity and satisfying Laplace's equation,
\begin{eqnarray}
\psi_{\rm m}=Ae^{-|{\bf k}|z}e^{{\rm i}\left(k_xx+k_yy+\omega t\right)},
\label{eq:vf}
\end{eqnarray}
$A$ is a constant and the subscript s indicates that the condition is to be applied at the surface.

The matching into a vacuum field is translated into a condition on the compressional Alfv\'en wave. Combining equations (\ref{eq:bc1_1}) and (\ref{eq:vf}), and writing the result in terms of dimensionless variables, we find, for axisymmetric modes,
\begin{eqnarray}
({\epsilon_{\perp}}^\prime-K\epsilon_{\perp})_{\rm s}=0,
\label{eq:perp}
\end{eqnarray}
where \,\, $\epsilon_{\perp}=Z\cos\alpha-U\sin\alpha$ \,\, is the component of the dimensionless displacement perpendicular to the direction of the magnetic field and $\alpha$ is the angle between the magnetic field direction and the x axis.
This boundary condition is applied sufficiently high in the atmosphere, in a region where $\epsilon_{\perp}^\prime / \epsilon_{\perp} \ll 1 / H$ (which implies that $(\sigma H/b)^2\varrho\ll 1$), with   
$H={\rm p}^\prime/{\rm p}$.

Next we consider two possibilities for the acoustic wave associated with the displacement parallel to the direction of the magnetic field. In some models this component of the solution is artificially reflected by imposing the condition
\begin{eqnarray}
 (\nabla\cdot{\mbox{\boldmath $\xi$}})_{\rm s}=0,
\label{eq:fr}
\end{eqnarray}
which, in terms of dimensionless variables becomes,
\begin{eqnarray}
 (Z'-{\rm i}KU)_{\rm s}=0.
\label{eq:tr}
\end{eqnarray}
In other models the same component of the solution was matched into the approximate analytic solution appropriate for an isothermal atmosphere. Starting from equations (\ref{eq:u})--(\ref{eq:g}) adapted for axisymmetric modes in an isothermal atmosphere, and considering the limit when the magnetic pressure is much larger than the gas pressure, we find that the dimensionless component of the solution parallel to the magnetic field direction, \,\, $\epsilon_{\parallel}=Z\sin\alpha+U\cos\alpha$, \,\, obeys approximately the equation
\begin{eqnarray}
{\epsilon_{\parallel}}^{\prime\prime}+\left(\frac{1}{H}-2{\rm i}\frac{K}{\tan\alpha}\right){\epsilon_{\parallel}}^\prime+\left(\frac{\sigma^2\varrho}{{\mathcal C}\gamma {\rm p}\sin^2\alpha}-\frac{K^2}{\tan^2\alpha}-{\rm i}\frac{K}{H\tan\alpha}\right)\epsilon_{\parallel}=0.
\label{eq:par}
\end{eqnarray}
The former equation accepts the solution
\begin{eqnarray}
\epsilon_{\parallel}=\frac{\tilde A}{{\rm p}^{1/2}}{\rm e}^{{\rm i}\left(K_a\eta +\phi\right)}
\label{eq:acou}
\end{eqnarray}
where $\tilde A$ is a constant, $\phi$ is a phase and
\begin{eqnarray}
K_a=\frac{K}{\tan\alpha}\pm \sqrt{\frac{\sigma^2\varrho}{{\mathcal C}\gamma {\rm p}\sin^2\alpha}-\frac{1}{4H^2}}.
\label{eq:K_a}
\end{eqnarray}
The numerical solution corresponding to the displacement parallel to the magnetic field direction is matched at the surface onto the analytical solution (\ref{eq:acou}), by imposing the condition,
\begin{eqnarray}
{\epsilon_{\parallel}}^{\prime}-\left(-\frac{1}{2H}+{\rm i}K_a\right)\epsilon_{\parallel}=0.
 \label{eq:bc2}
\end{eqnarray}
When the argument of the square root is positive, $K_a$ is real and the sign in equation (\ref{eq:K_a}) is chosen such as to make the solution correspond to an outwardly propagating wave. When the argument of the square root is negative, $K_a$ is complex and the sign in equation (\ref{eq:K_a}) is chosen such as to make the solution correspond to an evanescent wave.

\subsection*{Interior boundary condition}

In the interior, where the magnetic pressure becomes negligible when compared with the gas pressure, the magnetoacoustic wave decouples again into a wave which is essentially magnetic, a slow Alfv\'en wave, and a wave which is essentially acoustic. Here we follow the boundary condition applied by \citet{cunha00}. To leading order the acoustic component in the interior of the star is described by the system of equations (\ref{eq:u})--(\ref{eq:g}) with ${\bf B}$ set equal to zero, namely,   
\begin{eqnarray}
-\omega^2\rho u_{\rm a}=-{\rm i}|{\bf k}|p_1,
\end{eqnarray}
and
\begin{eqnarray}
-\omega^2\rho\xi_{z{\rm a}}=-\frac{\partial p_1}{\partial z}+g\rho_1.
\end{eqnarray}
Moreover, the displacement associated with the slow Alfv\'enic component is essentially along the horizontal direction (along the x axis for axisymmetric modes).
Following \citet{roberts83}, this component is assumed to dissipate in the interior and thus the corresponding numerical solution is matched onto the analytical asymptotic solution for an inwardly propagating slow Alfv\'en wave, derived in the JWKB approximation, namely, 
\begin{eqnarray}
%{\left(u_{\rm m}\right)}\sim & &\\ \nonumber
%& & \hspace{-2.3cm}D\rho^{-{1/4}}\exp{\left[{\rm i}\int_{0}^{z}{\left(\frac{\mu_0\rho\omega^2}{{B_z}^2}\right)^ {1/2}}{\rm d}z-{\rm i}\frac{kB_x}{B_z}z\right]}\exp\left[i\left(kx+\omega t\right)\right].
u_{\rm m}\sim D\rho^{-{1/4}}\exp{\left[{\rm i}\int_{0}^{z}{\left(\frac{\mu_0\rho\omega^2}{{B_z}^2}\right)^ {1/2}}{\rm d}z-{\rm i}\frac{kB_x}{B_z}z\right]}\exp\left[i\left(kx+\omega t\right)\right],
\label{eq:mas}
\end{eqnarray}
where $D$ is a (complex) constant and $u_m=u-({\rm i}|{\bf k}|p_1)/(\omega^2\rho)$ is the magnetic part of the horizontal displacement. In terms of dimensionless variables, the matching condition thus become,
\begin{eqnarray}
\frac{U_{\rm m}^\prime}{U_{\rm m}}={\rm i}\left[-\left(\frac{
\sigma^2\varrho}{b_z^2}\right)^{1/2}+K\frac{b_x}{b_z}\right],
\label{eq:upu}
\end{eqnarray}
with
\begin{eqnarray}
U_{\rm m}=U-{\rm i}\frac{{\mathcal C}K}{\sigma^2\varrho}\left(1-\frac{{\mathcal C}\gamma{\rm p}K^2}{\sigma^2\varrho}\right)^{-1}\left(Z{\rm p}^\prime +{\gamma{\rm p}} Z^{\prime}\right).
\end{eqnarray}

The boundary conditions defined above allow for the complete determination of the solution of the equations in the magnetic boundary layer (except for the amplitude which cannot be determined under the linear approximation). Thus, we may then use the numerical solution for the vertical component of the acoustic mode in the interior to match onto the corresponding asymptotic solution. Using equation (\ref{eq:xi}), that matching is translated to the condition (in dimensionless variables defined in the plane-parallel approximation),
\begin{eqnarray}
\frac{Z^\prime}{Z}\sim -{\mathcal K}\tan\delta,
\label{eq:zpz}
\end{eqnarray}
which is applied at the matching depth $\eta^*=1-R^*/R$, with ${\mathcal K}=R\kappa$. Equation (\ref{eq:zpz}) allows us to calculate the complex phase $\delta$ up to a factor of $\pi$. 
\end{appendix}
\bsp 

%%%%%%%%%%%%%%%%%%%%%%%%%%%%%%%%%%%%%%%%%%%%%%%%%%%%%%%%%%%%%%%%
\end{document}